\documentclass[times,10pt]{cm13article}
\usepackage{amsmath,amssymb,amsthm,mathrsfs,graphicx}
\usepackage{natbib,subfigure}

\usepackage{url}

\begin{document}

\begin{frontmatter}

\title{A goal-oriented reduced basis method for the wave equation in inverse analysis}           
\author[loc1]{K. C. Hoang\corref{cor1}}       
\ead{hoangk@cardiff.ac.uk}              
\author[loc1]{P. Kerfriden}                        
\author[loc1]{S. P. A. Bordas}                     
\cortext[cor1]{Corresponding author}
\address[loc1]{Institute of Mechanics and Advanced Materials, Cardiff University, UK}                 

\begin{abstract}
\noindent In this paper, we extend the reduced-basis methods developed earlier for wave equations to goal-oriented wave equations with affine parameter dependence. The essential new ingredient is the dual (or adjoint) problem and the use of its solution in a sampling procedure to pick up ``goal-orientedly'' parameter samples. First, we introduce the reduced-basis recipe --- Galerkin projection onto a space $Y_N$ spanned by the reduced basis functions which are constructed from the solutions of the governing partial differential equation at several selected points in parameter space. Second, we propose a new ``goal-oriented'' Proper Orthogonal Decomposition (POD)--Greedy sampling procedure to construct these associated basis functions. Third, based on the assumption of affine parameter dependence, we use the offline-online computational procedures developed earlier to split the computational procedure into offline and online stages. We verify the proposed computational procedure by applying it to a three-dimensional simulation dental implant problem. The good numerical results show that our proposed procedure performs better than the standard POD--Greedy procedure in terms of the accuracy of output functionals.

\end{abstract}


\end{frontmatter}

\section{Introduction}\label{sec1}

The design, optimization and control procedures of engineering problems often require several forms of performance measures or outputs --- such as displacements, heat fluxes or flowrates \cite{Grepl2005}. Generally, these outputs are functions of field variables such as displacements, temperature or velocities which are usually governed by a partial differential equation (PDE). The parameter or input will typically define a particular configuration of the problem. The relevant system behavior will thus be described by an implicit input-output relationship; evaluation of which requires the solution of the underlying parameter-PDE (or $\mu$PDE). We pursue the reduced-basis method \cite{Hoang2012, kerfriden2011local} which permits the \textit{efficient} and \textit{reliable} evaluation of this PDE-induced input-output relationship in \textit{real-time} and \textit{many queries} contexts.

The reduced-basis (RB) method was first introduced in the late 1970s for nonlinear analysis of structures and has been further investigated and developed more broadly \cite{PateraWebsite}. In particular, the RB method was well developed for various kinds and classes of parametrized PDEs: the eigenvalue problems, the coercive/non-coercive affine/non-affine linear/nonlinear elliptic PDEs, the coercive/non-coercive affine/non-affine linear/nonlinear parabolic
PDEs, the coercive affine linear hyperbolic PDEs, and several nonlinear problems such as Navier-Stokes equation, Burger's equation and Boussinesq equation \cite{PateraWebsite}. For the linear wave equation, the RB method and associated \textit{a posteriori} error estimation was developed successfully with some levels \cite{Tan2006, Huynh2011b, Hoang2012}; however, non of these works have focused on goal-oriented or dual problem of the wave equation.

In this work, we focus and improve significantly the output computation associated with the linear wave equation by proposing a new goal-oriented POD--Greedy algorithm. The paper is organized as follows. In Section \ref{sec2}, we introduce the necessary notation and state the problem. The RB approximation and the new goal-oriented POD--Greedy algorithm are discussed in Section \ref{sec3}. In Section \ref{sec4}, some numerical results of the dental implant problem \cite{Hoang2012} are presented to show the preeminence of the proposed algorithm. Finally, we provide some concluding remarks in Section \ref{sec5}.

\section{Problem Statement}\label{sec2}

\subsection{Abstract Formulation}\label{sec2_subsec1}

We consider a spatial domain $\Omega \in \mathbb{R}^3$ with a Lipschitz continuous boundary $\partial \Omega$. We denote the Dirichlet portion of the boundary by $\partial \Omega^D$. We then introduce the Hilbert spaces $Y^e \equiv (H^1_0(\Omega))^3$ and $X^e \equiv (L^2(\Omega))^3$, where $H^1(\Omega) = \{ v \, | \, v \in L^2(\Omega), \, \nabla v \in (L^2(\Omega))^3 \}$, $H^1_0(\Omega)=\{v \, | \, v \in H^1(\Omega), v| _{\partial \Omega^D}=0\}$ and $L^2(\Omega)$ is the space of square integrable functions over $\Omega$. The inner product and norm associated with $Y^e$ ($X^e$) are given by $(\cdot,\cdot)_{Y^e}$ $\left((\cdot,\cdot)_{X^e}\right)$ and $\|\cdot\|_{Y^e}=(\cdot,\cdot)_{Y^e}^{1/2}$ ($\|\cdot\|_{X^e}=(\cdot,\cdot)_{X^e}^{1/2}$), respectively; for example, $(w,v)_{Y^e} = \int_{\Omega} \frac{\partial w_i}{\partial x_j} \frac{\partial v_i}{\partial x_j} + w_i v_i$, $\forall w,v \in Y^e$ and $(w,v)_{X^e} = \int_{\Omega} w_i v_i$, $\forall w,v \in X^e$.

For time integration, we divide the time interval $[0,T]$ into $K$ subintervals of equal lengths $\Delta t=\frac{T}{K}$, and define $t^k=k\Delta t, \, 0 \le k \leq K$. We shall consider the Newmark's scheme with coefficients $(\varphi=\frac{1}{2}, \psi=\frac{1}{4})$ \cite{Hoang2012} for the time integration. Clearly, our results must be stable as $\Delta t \rightarrow 0$, $K \rightarrow \infty$.

We next define our parameter set $\mathcal{D} \in \mathbb{R}^P$, a typical point in which shall be denoted $\mu \equiv (\mu_1,\ldots,\mu_P)$. We then define the parametrized bilinear forms $a$ in $Y^e$, $a:Y^e \times Y^e \times \mathcal{D} \rightarrow \mathbb{R}$; $m,c,f,\ell$ are continuous bilinear and linear forms in $X^e$, $m: X^e \times X^e \rightarrow \mathbb{R}$, $c:X^e \times X^e \times \mathcal{D} \rightarrow \mathbb{R}$, $f: X^e \rightarrow \mathbb{R}$ and $\ell: X^e \rightarrow \mathbb{R}$.

The ``exact'' linear elasticity problem is stated follows: given a parameter $\mu \in \mathcal{D} \subset \mathbb{R}^{P}$, we evaluate the output of interest

\begin{equation}\label{eq:one}
s^e(\mu,t^k) = \ell (u^e(\mu,t^k)), \quad 0 \le k \le K,
\end{equation}

\noindent where the field variable $u^e(\mu,t^k) \in Y^e$ satisfies the weak form of the $\mu$-parametrized hyperbolic PDE \cite{Hoang2012}

\begin{multline}\label{eq:two}
m(u^e(\mu,t^{k+1}),v) + \frac{1}{2}\Delta t c(u^e(\mu,t^{k+1}),v;\mu) + \frac{1}{4} \Delta t^2 a(u^e(\mu,t^{k+1}),v;\mu) =
- m(u^e(\mu,t^{k-1}),v) + \frac{1}{2}\Delta t c(u^e(\mu,t^{k-1}),v;\mu)  \\  - \frac{1}{4} \Delta t^2 a(u^e(\mu,t^{k-1}),v;\mu)
+ 2m(u^e(\mu,t^{k}),v) - \frac{1}{2} \Delta t^2 a(u^e(\mu,t^{k}),v;\mu) + \Delta t^2 g^{eq}(t^k) f(v), \quad \forall v \in Y^e, 1 \leq k \leq K-1,
\end{multline}

\noindent with initial conditions $u^e(\mu,t^0)=0$, $\frac{\partial u^e(\mu,t^0)}{\partial t}=0$ and $g^{eq}(t^k) = \frac{1}{4} g(t^{k-1}) + \frac{1}{2} g(t^k) + \frac{1}{4} g(t^{k+1}), \quad 1 \le k \le K-1.$

We next introduce a reference finite element approximation space $Y \subset Y^e (\subset X^e)$ of dimension $\mathcal{N}$; we further define $X \equiv X^e$. Note that $Y$ and $X$ shall inherit the inner product and norm from $Y^e$ and $X^e$, respectively. Our ``true'' finite element approximation $u(\mu,t^k) \in Y$ to the ``exact'' problem is stated as:

\begin{multline}\label{eq:three}
m(u(\mu,t^{k+1}),v) + \frac{1}{2}\Delta t c(u(\mu,t^{k+1}),v;\mu) + \frac{1}{4} \Delta t^2 a(u(\mu,t^{k+1}),v;\mu) =
- m(u(\mu,t^{k-1}),v) + \frac{1}{2}\Delta t c(u(\mu,t^{k-1}),v;\mu)  \\  - \frac{1}{4} \Delta t^2 a(u(\mu,t^{k-1}),v;\mu)
+ 2m(u(\mu,t^{k}),v) - \frac{1}{2} \Delta t^2 a(u(\mu,t^{k}),v;\mu) + \Delta t^2 g^{eq}(t^k) f(v), \quad \forall v \in Y, 1 \leq k \leq K-1,
\end{multline}

\noindent with initial conditions $u(\mu,t^0)=0$, $\frac{\partial u(\mu,t^0)}{\partial t}=0$ and $g^{eq}(t^k)$ is defined as above. We then evaluate the output of interest

\begin{equation}\label{eq:four}
s(\mu,t^k) = \ell (u(\mu,t^k)), \quad 0 \le k \le K.
\end{equation}

The reduced basis approximation shall be built upon our reference finite element approximation, and the reduced basis error will thus be evaluated with respect to $u(\mu,t^k) \in Y$. Clearly, our methods must remain computationally efficient and stable as $\mathcal{N} \rightarrow \infty$.

We shall make the following assumptions. First, we assume that the bilinear forms $a(\cdot,\cdot;\mu)$ and $m(\cdot,\cdot;\mu)$ are continuous, coercive and symmetric \cite{Hoang2012}. Second, we require that all linear and bilinear forms are independent of time -- the system is thus linear time-invariant (LTI) \cite{Grepl2005}. And third, we shall assume that the bilinear forms $a$ and $c$ depend affinely on the parameter $\mu$ and can be expressed as

\begin{subequations}\label{eq:five}
\begin{equation}
a(w,v;\mu) = \displaystyle \sum_{q=1}^{Q_a} \Theta^q_a(\mu)a^q(w,v), \quad \forall w,v \in Y, \mu \in \mathcal{D}, \label{eq:five_sub1}
\end{equation}
\begin{equation}
c(w,v;\mu) = \displaystyle \sum_{q=1}^{Q_c} \Theta^q_c(\mu)c^q(w,v), \quad \forall w,v \in Y, \mu \in \mathcal{D}. \label{eq:five_sub2}
\end{equation}
\end{subequations}

To ensure rapid convergence of the reduced-basis output approximation we introduce a dual (or adjoint) problem which shall evolve backward in time \cite{Grepl2005}. Let $\tilde{t}=T-t$, then the dual solution $z(\mu,\tilde{t}^k)$ shall satisfies the following semi-discrete dual problem \cite{Bangerth2010}

\begin{multline}\label{eq:six}
m(z(\mu,\tilde{t}^{k+1}),v) + \frac{1}{2}\Delta \tilde{t} c(z(\mu,\tilde{t}^{k+1}),v;\mu) + \frac{1}{4} \Delta \tilde{t}^2 a(z(\mu,\tilde{t}^{k+1}),v;\mu) = - m(z(\mu,\tilde{t}^{k-1}),v) + \frac{1}{2}\Delta \tilde{t} c(z(\mu,\tilde{t}^{k-1}),v;\mu)  \\  - \frac{1}{4} \Delta \tilde{t}^2 a(z(\mu,\tilde{t}^{k-1}),v;\mu)
+ 2m(z(\mu,\tilde{t}^{k}),v) - \frac{1}{2} \Delta \tilde{t}^2 a(z(\mu,\tilde{t}^{k}),v;\mu) + \Delta \tilde{t}^2 \ell(v), \quad \forall v \in Y, 1 \leq k \leq \tilde{K}-1,
\end{multline}

\noindent with ``final'' conditions: $z(\mu,\tilde{t}^0)=0$, $\frac{\partial z(\mu,\tilde{t}^0)}{\partial \tilde{t}}=0$. The use of this dual equation will be clear in the next sections.

\subsection{Impulse Response}\label{sec2_subsec2}

In many dynamical systems, generally, the applied force to excite the system ($g(t^k)$ and $g^{eq}(t^k)$ in \eqref{eq:three}) is not known in advance (or \textit{a priori}) and thus we cannot solve \eqref{eq:three} for $u(\mu,t^k)$. In such situations, fortunately, we may appeal to the LTI hypothesis to justify an impulse approach as described now \cite{Grepl2005}. We note from the Duhamel's Principle that the solution of any LTI system can be written as the convolution of the impulse response with the control input: for any control input $g_{any}(t^k)$ (and hence $g^{eq}_{any}(t^k)$, $1 \le k \le K-1$), we can obtain $u_{any}(\mu,t^k)$, $1 \leq k \le K$ from:

\begin{equation}\label{eq:seven}
u_{any}(\mu,t^k)=\sum_{j=1}^{k} u_{unit}(\mu,t^{k-j+1}) g_{any}^{eq}(t^j), \quad 1 \le k \le K,
\end{equation}

\noindent where $u_{unit}(\mu,t^k)$ is the solution of \eqref{eq:three} for a unit impulse control input $g_{unit}(t^k)=\delta_{1k}$, $1 \le k \le K$. Therefore, it is sufficient to perform all computations related to FEM and RB approximations based on this impulse response \cite{Grepl2005}.

\section{Reduced Basis Approximation}\label{sec3}

\subsection{Reduced Basis Method}\label{sec3_subsec1}

We introduce the nested sample sets $S^{pr}_{N_{pr}}=\{\mu^{pr}_{1}\in \mathcal{D}, \ldots, \mu^{pr}_{N_{pr}}\in \mathcal{D}\}$, $1 \le N_{pr} \le N_{pr,\max}$, and $S^{du}_{N_{du}}=\{\mu^{du}_{1}\in \mathcal{D}, \ldots, \mu^{pr}_{N_{du}}\in \mathcal{D}\}$, $1 \le N_{du} \le N_{du,\max}$. Here, $N_{pr}$ and $N_{du}$ are the dimensions of the reduced basis space for the primal and dual variables, respectively; in general, $S^{pr}_{N_{pr}} \neq S^{du}_{N_{du}}$ and in fact $N_{pr} \neq N_{du}$. We then define the associated nested Lagrangian reduced basis spaces

\begin{subequations}\label{eq:eight}
\begin{equation}
Y^{pr}_{N_{pr}} = {\rm span} \{\zeta^{pr}_{n}, \, 1 \le n \le N_{pr}\}, \quad 1 \le N_{pr} \le N_{pr,\max}, \label{eq:eight_sub1}
\end{equation}
\begin{equation}
Y^{du}_{N_{du}} = {\rm span} \{\zeta^{du}_{n}, \, 1 \le n \le N_{du}\}, \quad 1 \le N_{du} \le N_{du,\max}.
\label{eq:eight_sub2}
\end{equation}
\end{subequations}

The reduced basis approximation $u_N(\mu,t^k)$ to $u(\mu,t^k)$ is then obtained by a standard Galerkin projection: given $\mu \in \mathcal{D}$, $u_N(\mu,t^k) \in Y^{pr}_{N_{pr}}$ satisfies

\begin{multline}\label{eq:nine}
m(u_N(\mu,t^{k+1}),v) + \frac{1}{2}\Delta t c(u_N(\mu,t^{k+1}),v;\mu) + \frac{1}{4} \Delta t^2 a(u_N(\mu,t^{k+1}),v;\mu) =
- m(u_N(\mu,t^{k-1}),v) + \frac{1}{2}\Delta t c(u_N(\mu,t^{k-1}),v;\mu)  \\  - \frac{1}{4} \Delta t^2 a(u_N(\mu,t^{k-1}),v;\mu) + 2m(u_N(\mu,t^{k}),v) - \frac{1}{2} \Delta t^2 a(u_N(\mu,t^{k}),v;\mu) + \Delta t^2 g^{eq}(t^k) f(v), \quad \forall v \in Y^{pr}_{N_{pr}}, 1 \leq k \leq K-1,
\end{multline}

\noindent with initial conditions: $u_N(\mu,t^0)=0$, $\frac{\partial u_N(\mu,t^0)}{\partial t}=0$. Similarly, the reduced basis approximation $z_N(\mu,\tilde{t}^k) \in Y^{du}_{N_{du}}$ to $z(\mu,\tilde{t}^k)$ is obtained from

\begin{multline}\label{eq:ten}
m(z_N(\mu,\tilde{t}^{k+1}),v) + \frac{1}{2}\Delta \tilde{t} c(z_N(\mu,\tilde{t}^{k+1}),v;\mu) + \frac{1}{4} \Delta \tilde{t}^2 a(z_N(\mu,\tilde{t}^{k+1}),v;\mu) = - m(z_N(\mu,\tilde{t}^{k-1}),v) + \frac{1}{2}\Delta \tilde{t} c(z_N(\mu,\tilde{t}^{k-1}),v;\mu)  \\  - \frac{1}{4} \Delta \tilde{t}^2 a(z_N(\mu,\tilde{t}^{k-1}),v;\mu)
+ 2m(z_N(\mu,\tilde{t}^{k}),v) - \frac{1}{2} \Delta \tilde{t}^2 a(z_N(\mu,\tilde{t}^{k}),v;\mu) + \Delta \tilde{t}^2 \ell(v), \quad \forall v \in Y^{du}_{N_{du}}, 1 \leq k \leq \tilde{K}-1,
\end{multline}

\noindent with ``final'' conditions: $z_N(\mu,\tilde{t}^0)=0$, $\frac{\partial z_N(\mu,\tilde{t}^0)}{\partial \tilde{t}}=0$. Finally, we evaluate the output estimate, $s_N(\mu,t^k)$, from \cite{Grepl2005}

\begin{equation}\label{eq:eleven}
s_N(\mu,t^k) = \ell(u_N(\mu,t^k)) + \mathfrak{R}(z_N(\mu,\tilde{t}^k);\mu,t^k), \quad 1 \le k \le K,
\end{equation}

\noindent where $\mathfrak{R}(z_N(\mu,\tilde{t}^k);\mu,t^k) = \Delta t \displaystyle \sum_{k'=1}^{k} \mathcal{R}^{pr}(z_N(\mu,\tilde{t}^{k+1-k'});\, \mu,t^{k'})$. Here, we note that the terms $\mathcal{R}^{pr}(v;\mu,t^k)$ and $\mathcal{R}^{du}(v;\mu,\tilde{t}^k)$ are the primal and dual residual associated with the RB equations \eqref{eq:nine} and \eqref{eq:ten}, respectively

\begin{subequations}\label{eq:twelve}
\begin{multline}
\mathcal{R}^{pr}(v;\mu,t^k) = g^{eq}(t^k)f(v) - \displaystyle \frac{1}{\Delta t^2} \left( m(u_N(\mu,t^{k+1}),v) - 2m(u_N(\mu,t^{k}),v) + m(u_N(\mu,t^{k-1}),v) \right) \\
- \displaystyle \frac{1}{\Delta t} \left( \frac{1}{2} c(u_N(\mu,t^{k+1}),v;\mu) - \frac{1}{2} c(u_N(\mu,t^{k-1}),v;\mu) \right) - \left( \frac{1}{4} a(u_N(\mu,t^{k+1}),v;\mu) + \frac{1}{2} a(u_N(\mu,t^{k}),v;\mu) + \frac{1}{4} a(u_N(\mu,t^{k-1}),v;\mu) \right),
\label{eq:twelve_sub1}
\end{multline}
\begin{multline}
\mathcal{R}^{du}(v;\mu,\tilde{t}^k) = \ell(v) - \displaystyle \frac{1}{\Delta \tilde{t}^2} \left( m(z_N(\mu,\tilde{t}^{k+1}),v) - 2m(z_N(\mu,\tilde{t}^{k}),v) + m(z_N(\mu,\tilde{t}^{k-1}),v) \right) \\
- \displaystyle \frac{1}{\Delta \tilde{t}} \left( \frac{1}{2} c(z_N(\mu,\tilde{t}^{k+1}),v;\mu) - \frac{1}{2} c(z_N(\mu,\tilde{t}^{k-1}),v;\mu) \right) - \left( \frac{1}{4} a(z_N(\mu,\tilde{t}^{k+1}),v;\mu) + \frac{1}{2} a(z_N(\mu,\tilde{t}^{k}),v;\mu) + \frac{1}{4} a(z_N(\mu,\tilde{t}^{k-1}),v;\mu) \right),
\label{eq:twelve_sub2}
\end{multline}
\end{subequations}

\noindent $\forall v \in Y, 1 \le k \le K$. Note that here $N \equiv (N_{pr}, N_{du})$.

\subsection{Computational Procedure}\label{sec3_subsec2}

The computational procedure for the primal and dual RB equations can be developed completely similar to that in our previous work \cite{Hoang2012} or in \cite{Grepl2005}. That is, it can be decomposed into two stages: offline and online stages thanks to the affine decomposition \eqref{eq:five}. The interested readers can refer to \cite{Hoang2012, Grepl2005} for more details.

\subsection{POD--Greedy Sampling Procedure}\label{sec3_subsec3}

In this section, we present the key contribution of this work, namely, the ``goal-oriented'' POD--Greedy sampling procedure. The standard POD--Greedy algorithm which has been used widely in the RB context for time-dependent problems \cite{Haasdonk2008, Huynh2011b, Nguyen2009} and our proposed ``goal-oriented'' POD--Greedy algorithm are presented simultaneously in the following table

\begin{table}[h!]
\begin{center}
  {\begin{tabular}{|l|l|}
  \hline
    Set $Y^{pr}_{N_{pr}}=0$   &   Set $Y^{pr}_{N_{pr}}=0$   \\
    Set $\mu^{pr}_{*}=\mu^{pr}_{0}$   &   Set $\mu^{pr}_{*}=\mu^{pr}_{0}$   \\
    While $N_{pr} \le N_{pr,\max}$   &   While $N_{pr} \le N_{pr,\max}$   \\
    \qquad $\mathcal{W}=\left\{e_{proj}(\mu^{pr}_{*},t^k), \, 0 \le k \le K \right\}$;   &
    \qquad $\mathcal{W}=\left\{e_{proj}(\mu^{pr}_{*},t^k), \, 0 \le k \le K \right\}$;   \\
    \qquad $Y^{pr}_{N_{pr}+M} \longleftarrow Y^{pr}_{N_{pr}} \bigoplus POD(\mathcal{W},M)$;   &
    \qquad $Y^{pr}_{N_{pr}+M} \longleftarrow Y^{pr}_{N_{pr}} \bigoplus POD(\mathcal{W},M)$;   \\
    \qquad $N_{pr} \longleftarrow N_{pr} + M$;   &   \qquad $N_{pr} \longleftarrow N_{pr} + M$;  \\
    \qquad $\mu^{pr}_{*}= \arg \max_{\mu \in \Xi_{train}} \left\{ \displaystyle \frac{ \sqrt{\sum_{k=1}^{K} \|\mathcal{R}^{pr}(v;\mu,t^k)\|^2_{Y'} } }{\sqrt{\sum_{k=1}^{K}\|u_N(\mu,t^k)\|^2_Y } } \right\} $;   &
    \qquad $\mu^{pr}_{*}= \arg \max_{\mu \in \Xi_{train}} \left\{ \displaystyle \frac{ \sqrt{\sum_{k=1}^{K} \mathfrak{R}^2(z_N(\mu,\tilde{t}^k);\mu,t^k) } }{ \sqrt{\sum_{k=1}^{K} s_N^2(\mu,t^k) } } \right\} $;   \\
    \qquad $S^* \longleftarrow S^* \bigcup \left\{ \mu^{pr}_* \right\} $;   &
    \qquad $S^* \longleftarrow S^* \bigcup \left\{ \mu^{pr}_* \right\} $;   \\
    end.    &   end.\\
  \hline
\end{tabular}
\caption{(Left) Standard POD--Greedy sampling algorithm and (Right) our proposed ``goal-oriented'' POD--Greedy sampling algorithm.}
\label{tab2}}
\end{center}
\end{table}

The main difference between our proposed ``goal-oriented'' POD--Greedy algorithm and the standard one is that we somehow try to minimize the error indicator of the output functional ($s_N(\mu,t^k)$) rather than minimize an error indicator of the field variable ($u_N(\mu,t^k)$) as in the standard POD--Greedy algorithm. By this way, we expect to improve the accuracy (or convergent rate) of the RB output functional approximation; but contrarily, we might lose the rapid convergent rate of the field variable as in the standard POD--Greedy algorithm.

\section{Numerical Results}\label{sec4}

\begin{figure}[h!]
 \centering
 \subfigure[]{ \includegraphics[height=4.2cm]{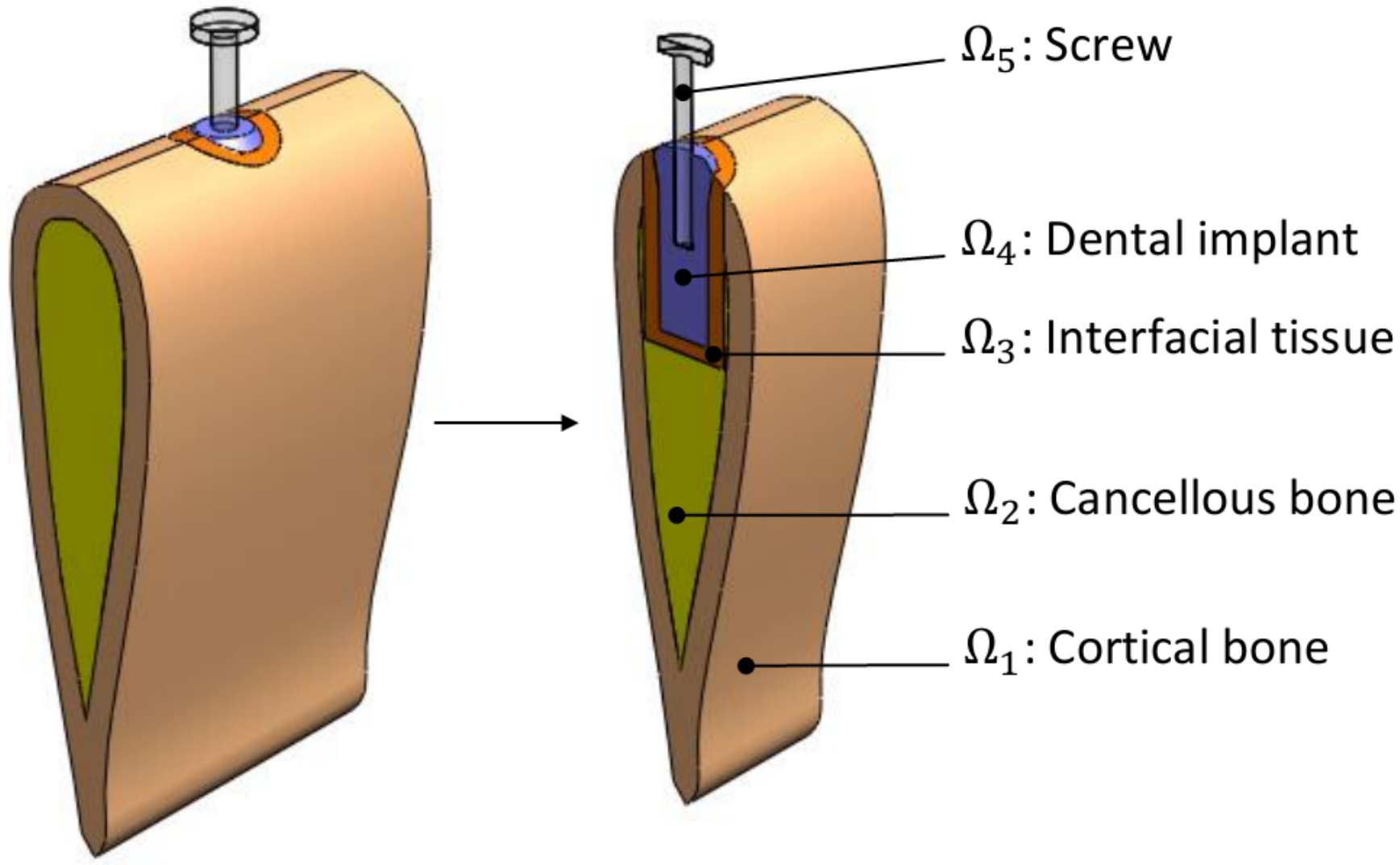}
 \label{fig1a}}
 \subfigure[]{ \includegraphics[height=4.2cm]{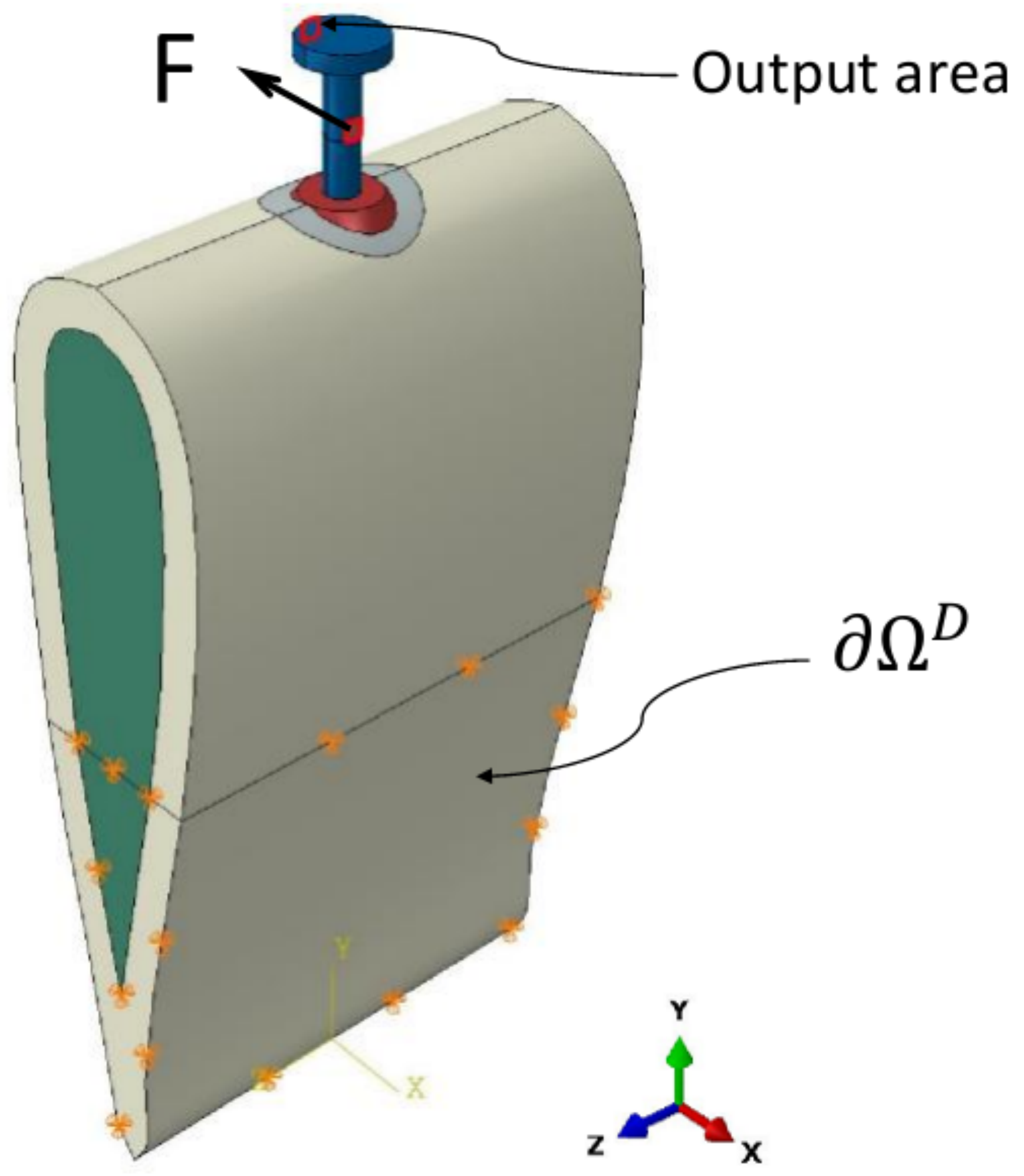}
 \label{fig1b}}
 \caption{(a) The 3d simplified FEM model with sectional view and (b) output area, applied load and boundary condition.}
 \label{fig1}
\end{figure}

In this section, we consider the dental implant problem \cite{Hoang2012} to verify the behavior of the proposed goal-oriented POD-Greedy algorithm. We consider a simplified 3D dental implant-bone model in Fig.\ref{fig1a}. The geometry of the simplified dental implant-bone model is constructed by using SolidWorks 2010. The physical domain $\Omega$ consists of five regions: the outermost cortical bone $\Omega_1$, the cancellous bone $\Omega_2$, the interfacial tissue $\Omega_3$, the dental implant $\Omega_4$ and the stainless steel screw $\Omega_5$. The 3D simplified model is then meshed and analyzed in the software ABAQUS/CAE version 6.10-1. A dynamic force opposite to the $x-$direction is then applied to the body of the screw as shown in Fig.\ref{fig1b}. The output of interest is defined as the average displacement responses of an area on the head of the screw (Fig.\ref{fig1b}). The Dirichlet boundary condition $(\partial \Omega^D)$ is specified in the bottom-half of the simplified model as illustrated in Fig.\ref{fig1b}. The finite element mesh consists of 9479 nodes and 50388 four-node tetrahedral solid elements. The coinciding nodes of the contact surfaces between different regions (the regions $\Omega_1$, $\Omega_2$, $\Omega_3$, $\Omega_4$, $\Omega_5$) are assumed to be rigidly fixed, i.e. the displacements in the $x-$, $y-$ and $z-$directions are all set to be the same for the same coinciding nodes.

We assume that the regions $\Omega_i, 1 \leq i \le 5$, of the simplified model are homogeneous and isotropic. The material properties: the Young's moduli, Poisson's ratios and densities of these regions are presented in Table \ref{tab1} \cite{Wang2010}. As similar to \cite{Hoang2012}, we still use Rayleigh damping with stiffness-proportional damping coefficient $\beta_i$, $1 \le i \le 5$ (Table \ref{tab1}) such that $\textbf{C}_i = \beta_i \textbf{A}_i, \, 1 \le i \le 5$, where $\textbf{C}_i$ and $\textbf{A}_i$ are the FEM damping and stiffness matrices of each region, respectively. We also note in Table \ref{tab1} that ($E_3$,$\beta_3$) are our sole parameters.

\begin{table}[h!]
\begin{center}
  {\begin{tabular}{|c|l|r|l|r|r|}
  \hline
 Domain & Layers & E (Pa) & $\nu$ & $\rho (\rm g/mm^3)$ & $\beta$ \\
  \hline
 $\Omega_1$ & Cortical bone         & $2.3162 \times 10^{10}$ & 0.371  & $1.8601 \times 10^{-3}$ & $3.38 \times 10^{-6}$ \\
 $\Omega_2$ & Cancellous bone       & $8.2345 \times 10^{8}$  & 0.3136 & $7.1195 \times 10^{-4}$ & $6.76 \times 10^{-6}$ \\
 $\Omega_3$ &  Tissue                &      E                  & 0.3155 & $1.055  \times 10^{-3}$ &         $\beta$ \\
 $\Omega_4$ &  Titan implant         & $1.05  \times 10^{11}$  & 0.32   & $4.52   \times 10^{-3}$ & $5.1791 \times 10^{-10}$ \\
 $\Omega_5$ &  Stainless steel screw & $1.93  \times 10^{11}$  & 0.305  & $8.027  \times 10^{-3}$ & $2.5685 \times 10^{-8}$ \\
   \hline
\end{tabular}
\caption{Material properties of the dental implant-bone structure.}
\label{tab1}}
\end{center}
\end{table}

\begin{figure}[h!]
 \centering
 \subfigure[]{ \includegraphics[height=4.2cm]{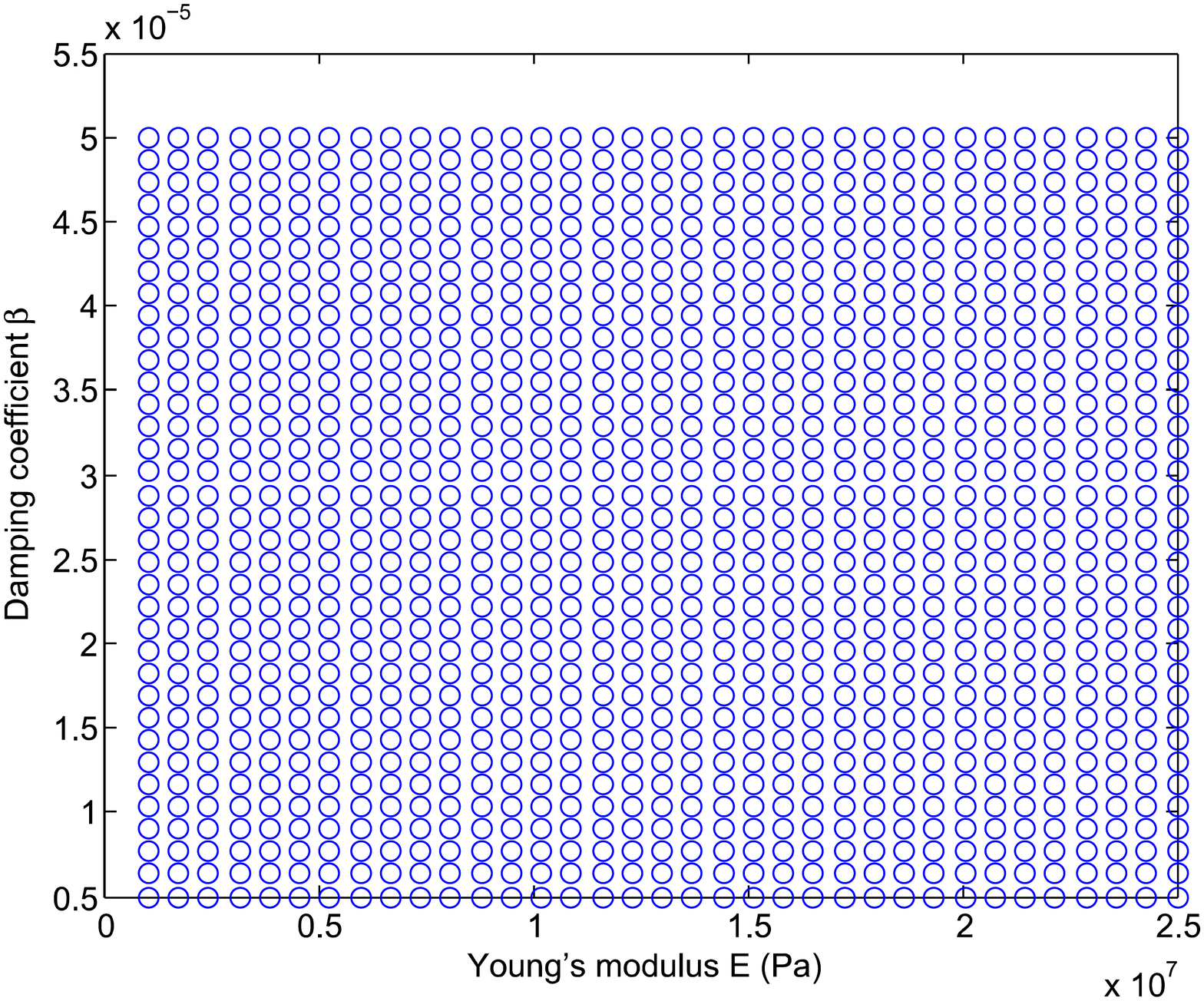}
 \label{fig2a}}
 \subfigure[]{ \includegraphics[height=4.2cm]{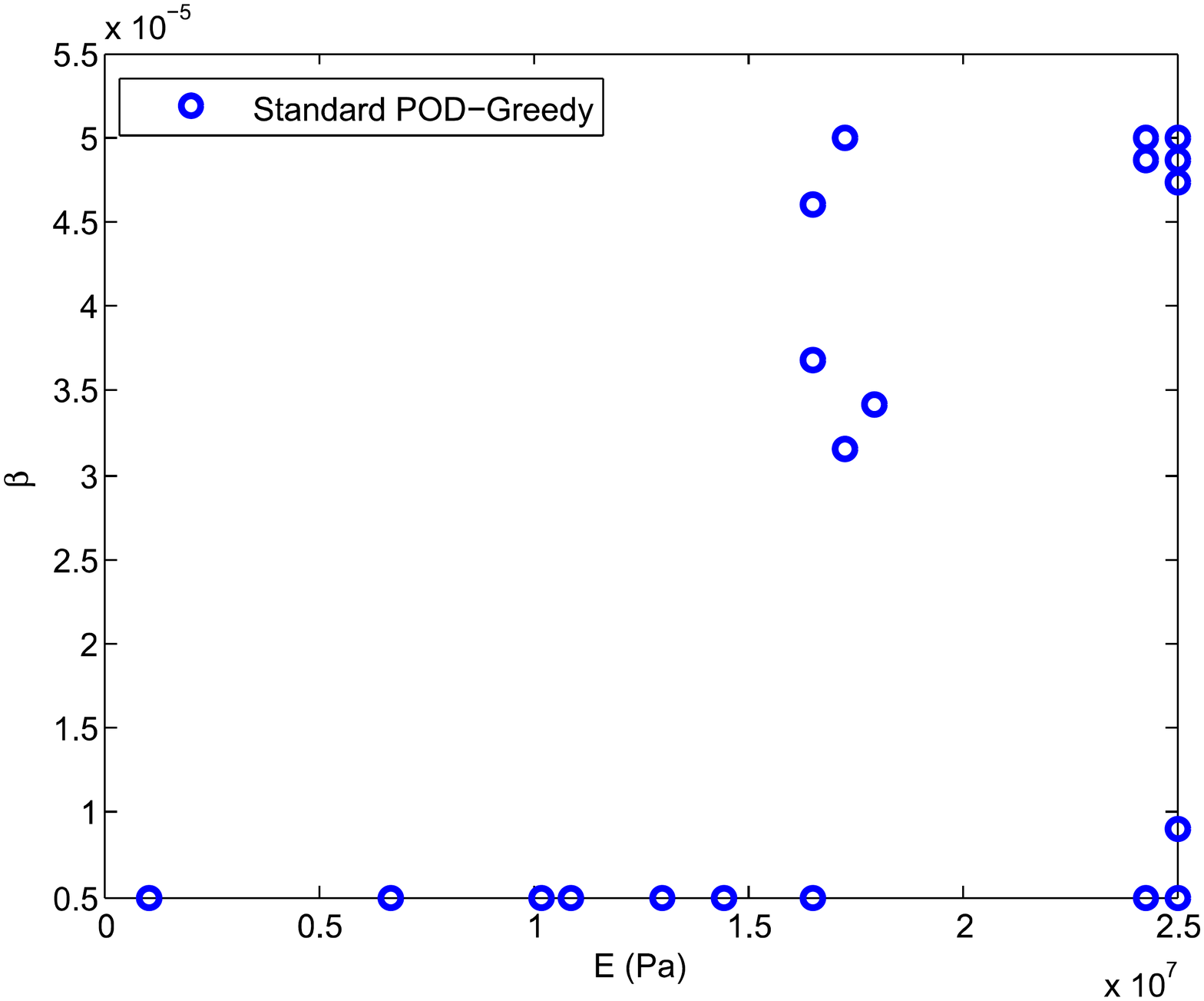}
 \label{fig2b}}
 \subfigure[]{ \includegraphics[height=4.2cm]{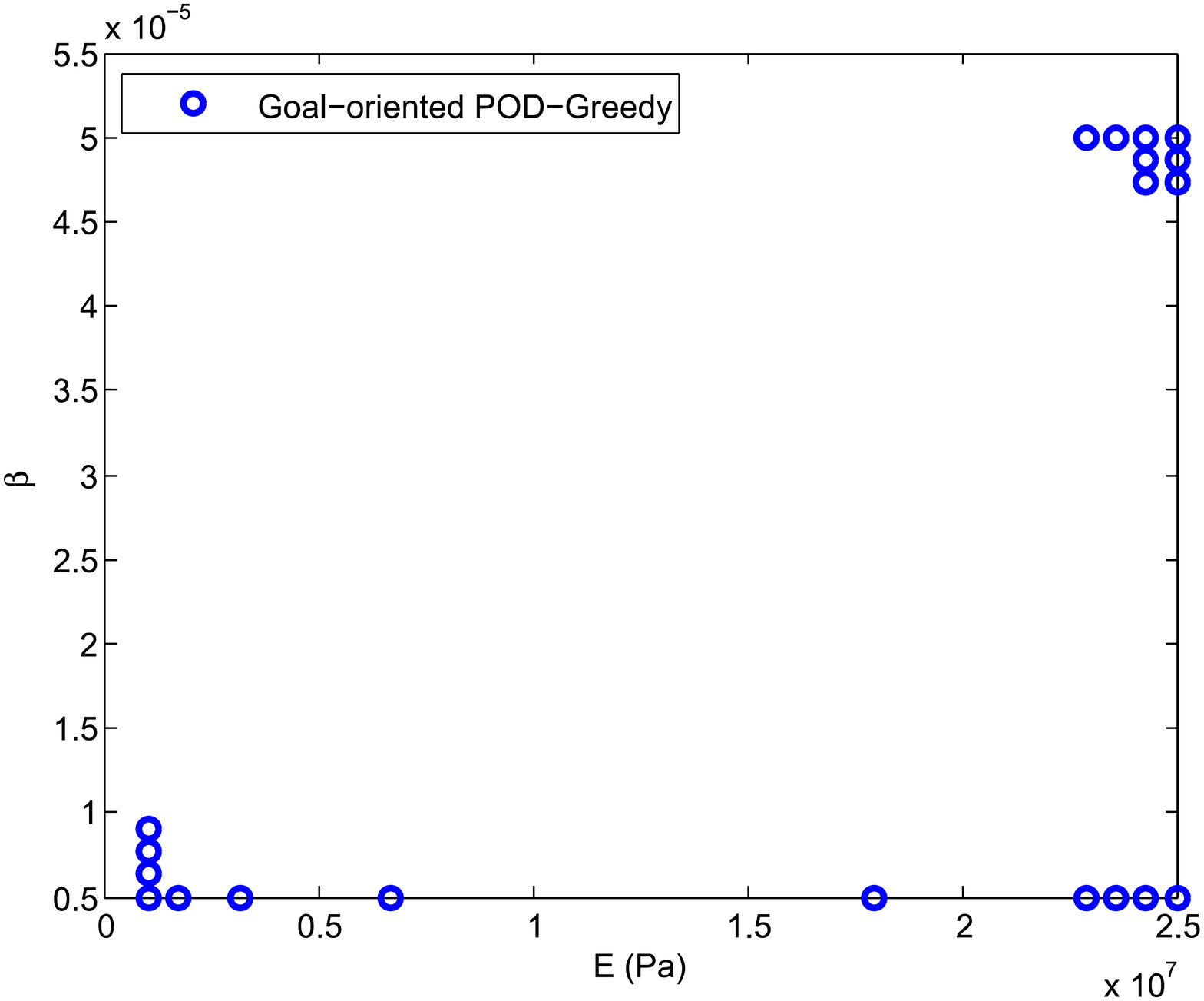}
 \label{fig2c}}
 \caption{(a) The $\Xi_{train}$ samples set. Distribution of sampling points by (b) standard POD--Greedy sampling algorithm and (c) goal-oriented POD--Greedy sampling algorithm. }
 \label{fig2}
\end{figure}

\begin{figure}[h!]
 \centering
 \subfigure[]{ \includegraphics[height=4.2cm]{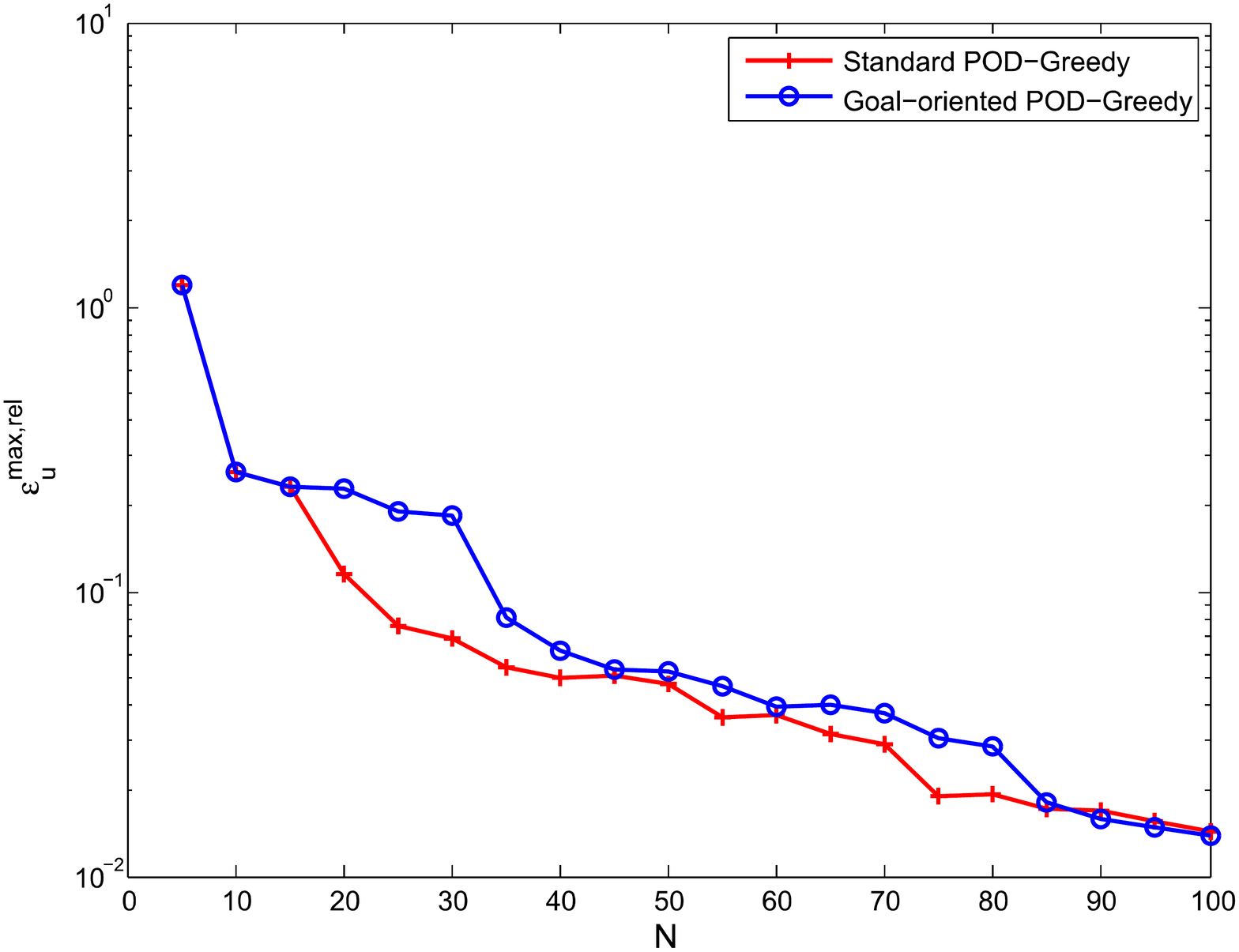}
 \label{fig3a}}
 \subfigure[]{ \includegraphics[height=4.2cm]{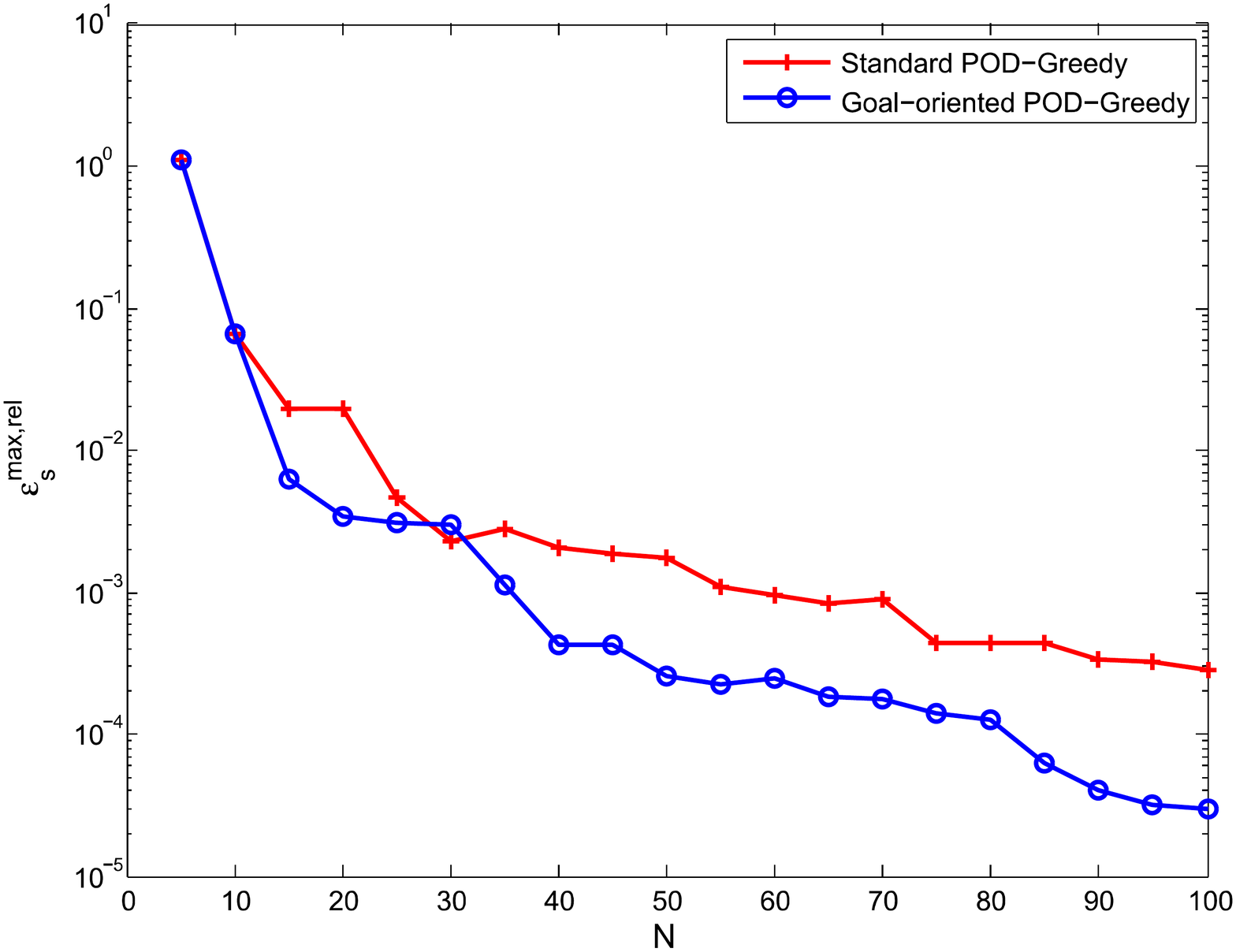}
 \label{fig3b}}
 \caption{Maximum relative RB error of (a) the solution and (b) the output by the two algorithms.}
 \label{fig2}
\end{figure}

We consider the FE space of dimension $\mathcal{N}=26343$. For time integration, $T=1 \times 10^{-3}$s, $\Delta t = \Delta \tilde{t}=2 \times 10^{-6}$s, $\tilde{K}=K=\frac{T}{\Delta t}=500$. The input parameter $\mu \equiv (E,\beta) \in \mathcal{D}$, where the parameter domain $\mathcal{D} \equiv [1 \times 10^6, 25 \times 10^6]{\rm Pa} \times [5 \times 10^{-6}, 5 \times 10^{-5}] \subset \mathbb{R}^{P=2}$. (Note that this parameter domain is nearly two times larger than that of \cite{Hoang2012}.) As shown in Fig.\ref{fig2a}, a sample set $\Xi_{train}$ is created by a uniform distribution over $\mathcal{D}$ with $n_{train}=1225$ samples. We implement both the standard and goal-oriented POD--Greedy algorithms for the primal equations \eqref{eq:three}, \eqref{eq:four}. In order to perform the goal-oriented POD--Greedy algorithm, note that we need the dual solution $z_N(\mu,\tilde{t}^k)$ for the $\mathfrak{R}$ term. Thus we use the standard POD--Greedy algorithm to build $Y^{du}_{N_{du}}$ in \eqref{eq:eight_sub2}, and then use $N_{du}=60$ basis functions for all computations related to this $\mathfrak{R}$ term. The distribution of sampling points by the standard and goal-oriented POD--Greedy algorithms are shown in Fig.\ref{fig2b} and Fig.\ref{fig2c}, respectively. Finally, we show as a function of $N_{pr}$: $\epsilon^{{\rm max,rel}}_u$ is the maximum over $\Xi_{train}$ of $\epsilon_u(\mu,t^K)$ and $\epsilon^{{\rm max,rel}}_s$ is the maximum over $\Xi_{train}$ of $\epsilon_s(\mu,t^K)$ in Fig.\ref{fig3a} and Fig.\ref{fig3b}, respectively\footnote{Note that the relative RB error is defined as: $\epsilon_u(\mu,t^K)=\displaystyle \frac{\sqrt{\sum_{k=1}^{K} \|u(\mu,t^k)-u_N(\mu,t^k)\|^2_Y}}{\sqrt{\sum_{k=1}^{K} \|u_N(\mu,t^k)\|^2_Y}}$; and $\epsilon_s(\mu,t^K)=\displaystyle \frac{\sqrt{\sum_{k=1}^{K} (s(\mu,t^k)-s_N(\mu,t^k))^2}}{\sqrt{\sum_{k=1}^{K} s^2_N(\mu,t^k)}}$.}. As observed, we see that the goal-oriented POD--Greedy algorithm improves significantly the convergent rate of the output while sacrificing a bit that of the solution (or field variable).

\section{Conclusions}\label{sec5}

A new ``goal-oriented'' POD--Greedy sampling algorithm was proposed. The proposed algorithm makes use of the primal residual of the dual solution rather than the dual norm of primal residual as error indicator in the standard POD--Greedy algorithm. The proposed algorithm is verified by investigating a 3D dental implant problem in the time domain. In comparison with the standard algorithm, we conclude that our proposed algorithm performs much better -- in terms of output's accuracy, and a little worse -- in terms of solution's accuracy.

%

\section*{Acknowledgements}

We are sincerely grateful for the financial support of the European Research Council Starting Independent Research
Grant for the project ERC No. 279578.

\bibliographystyle{cm13num}     
\bibliography{myrefs_ACME13}

\end{document}